\documentclass{emulateapj}

\usepackage{amsmath}
\usepackage{amssymb}
\usepackage{graphicx}
\usepackage{appendix}

\newcommand{\lSect}[1]{{\label{sec:#1}}}

\newcommand{\Msun}{\ensuremath{\mathrm{M}_\odot}}

\newcommand{\BE}{\ensuremath{B\!E}}

\newcommand{\E}[1]{\ensuremath{ \times 10^{#1}}}

\newcommand{\ep}{\epsilon}

\begin{document}

\title{Very Low Energy Supernovae from Neutrino Mass Loss}

\author{Elizabeth Lovegrove\altaffilmark{1} and S. E. Woosley\altaffilmark{1}}

\altaffiltext{1}{Department of Astronomy and Astrophysics, University
 of California, Santa Cruz, CA 95064; woosley@ucolick.org}

\begin{abstract}
The continuing difficulty of achieving a reliable explosion in
simulations of core-collapse supernovae, especially for more massive
stars, has led to speculation concerning the observable transients
that might be produced if such a supernova fails.  Even if a prompt
outgoing shock fails to form in a collapsing presupernova star, one
must still consider the hydrodynamic response of the star to the
abrupt loss of mass via neutrinos as the core forms a protoneutron
star. Following a suggestion by \citet{Nad80}, we calculate the
hydrodynamical responses of typical supernova progenitor stars to the
rapid loss of approximately 0.2 to 0.5 \Msun \ of gravitational mass
from their centers. In a red supergiant star, a very weak supernova
with total kinetic energy $\sim 10^{47}$ erg results. The binding
energy of a large fraction of the hydrogen envelope before the
explosion is of the same order and, depending upon assumptions
regarding the neutrino loss rates, most of it is ejected. Ejection
speeds are $\sim 100$ km s$^{-1}$ and luminosities $\sim 10^{39}$ erg
s$^{-1}$ are maintained for about a year. A significant part of the
energy comes from the recombination of hydrogen. The color of the
explosion is extremely red and the events bear some similarity to
``luminous red novae,'' but have much lower speeds.

\end{abstract}


\section{INTRODUCTION}
\lSect{intro}

Black holes are expected to form in a significant fraction of massive
star deaths \citep[e.g.,][]{Oco11,Ugl12} and it seems possible that in at
least some of these cases the black hole will form without generating an
outgoing shock. What observable signal would accompany such an event?
Would they all be ``un-novae'' \citep{Koc08}, stars that just
 disappear from view, or will some sort of transient display
announce the birth of any black hole? 

If the progenitor star was rotating, then a variety of energetic
transients might be possible depending on the distribution of angular
momentum, ranging from common gamma-ray bursts by the collapsar
mechanism \citep{Woo93,Mac99} to long duration x-ray
\citep{Mac01,Li03} and gamma-ray \citep{Qua12,Woo12} transients. But
what if the star were rotating very slowly or not at all? Pulsations
\citep{Woo07} or acoustic energy transport \citep{Qua12} may lead to
envelope ejection, but these either occur in stars that are uncommonly
massive or depend on details of gravity wave propagation that are
still to be worked out.

Here we follow on a suggestion by \citet{Nad80} which provides a
simple mechanism for ejecting at least some mass in supernova
progenitors with very weakly bound envelopes, i.e., red
supergiants. In all but extremely rare super-massive stars, iron core
collapse leads to protoneutron star formation \citep{Oco11}. The
protoneutron star must then emit a considerable fraction of its
mass-energy as neutrinos before either settling down as a cold
degenerate object or contracting inside its event horizon and becoming
a black hole. This is true even if the protoneutron star continues to
accrete substantial mass while contracting. In fact, in that case, the
total energy emitted should be close to the binding energy of a
neutron star with its maximum stable gravitational mass, which we
allow to vary in the range 2.0 - 2.5 \Msun in our model.

As frequently noted, this huge energy loss also results in a decrease
in the gravitational mass of the compact object
\citep{Lat89,Lat01}. This mass loss is of order
\begin{equation}
BE \approx 0.084 \left(\frac{M_G}{\Msun}\right)^2 \ \Msun \label{eq:BE}
\end{equation}
where $M_G$ is the gravitational mass of the cold neutron star. This
implies a mass loss of approximately 0.2 - 0.5 \Msun \ over a period
of several seconds, far shorter than the sound crossing time for the
helium and heavy element core. This ``mass'' streams out through the
presupernova star at the speed of light without interacting
appreciably. To the remaining star it simply appears that the
gravitational potential of the core has abruptly decreased. In
response, the star begins to expand. In this paper we simulate
this mass loss and the resulting expansion to determine whether it is
capable of creating a shock with sufficient energy to reach the outer
layers of the star and, if it does, to eject mass and form a visible
transient. It is important to note that this expansion precedes the
loss in support pressure caused by the collapse of the core. That
rarefaction passes out through the star on a slower, free-fall time
scale.

\section{Computational Procedure}

We began with two presupernova models calculated using the KEPLER
stellar evolution code. KEPLER is a Lagrangian 1D implicit
hydrodynamics code with the appropriate nuclear physics, mass loss,
opacities, and equations of state for studying massive star evolution
\citep{Wea78,Woo02}. It is not Courant-limited, allowing it to follow
a star from its formation to the collapse of its iron core. However,
because KEPLER uses a Lagrangian grid, it cannot easily implement an
absorbing inner boundary condition. Consequently, it cannot accurately
follow the infall of the star once a compact object has formed at its
center.  Once the presupernova models calculated by KEPLER reached
supersonic collapse speeds in their cores, they were thus mapped into
CASTRO\footnote{Code version as of May 2012.} for further study. CASTRO is a multidimensional Eulerian
radiation-hydrodynamics code with adaptive mesh refinement, stellar
equations of state, nuclear reaction networks and self-gravity
\citep{castro}. The calculations were done in one dimension with
reactions turned off, and the central object modeled as a point of
variable mass. An Eulerian grid with constant mesh and spherical
geometry was employed. After the collapse had proceeded long enough in
CASTRO to allow shock propagation to the base of the hydrogen
envelope, the results were then mapped back into KEPLER again and the
final hydrodynamics and a light curve were calculated.

\subsection{Central Object Modeling}

Hydrodynamics and radiation transport in the region immediately
surrounding a protoneutron star or black hole is complex and difficult
to model. A realistic simulation of the neutrino transport alone is
well beyond the scope of this paper and the outcome would depend upon
the chosen neutron star equation of state and the dimensionality of
the calculation. Since we are only interested in the temporal
evolution of the central gravitational potential, however, reasonably
good results can be obtained from parametrized calculations with a
generic, qualitative description of the neutrino energy loss.  The
inner boundary of our simulation is placed, in CASTRO, at the outer
edge of the pre-collapse iron core. Matter interior to this boundary
is treated as a point having variable gravitational mass. Initially
this mass is equal to the baryonic mass of the interior matter, i.e.,
the iron core mass. As time passes that gravitational mass decreases
due to neutrino emission and, if no mass were added, it would
eventually decrease to the known cold neutron star value for a given
baryonic mass and equation of state \citep{Pra97}. However, appreciable mass does accrete, leading
in the more interesting cases to the formation of a black hole. During
the accretion phase neutrinos continue to carry away effective
mass. Since we neglect rotation, we assume that, after the black hole
forms, matter and energy flow into the black hole without any more
emission.

An upper limit to the mass lost is given by the gravitational binding
energy of the maximum stable neutron star mass, the
``Tolman-Oppenheimer-Volkoff (TOV) limit'' \citep{Opp39}. Current best
limits place this parameter in the range 2.0 to 2.5 \Msun of gravitational mass
\ \citep{Akm98,Dem10}. The characteristic time scale for this loss,
which is assumed to occur exponentially, is given by the cooling
timescale parameter $\tau_c$, taken here to be 3 seconds
\citep[e.g.][]{Bur87}. For this simple case, the mass loss rate is
\begin{equation}
\dot M_{\rm G} =  \frac{\BE}{\tau_c} e^{-t/\tau_c} \label{eq:max}
\end{equation} 
This equation gives an upper bound on the mass loss and therefore on
the strength of the transient produced. This model is referred to as
the ``maximum mass loss'' case.

Somewhat more realistically, we can take into account the time for the
critical mass to accrete and follow the neutrino energy losses that
occur during that time. For each bit of mass that accretes, some
fraction of its gravitational binding energy is radiated away
promptly, but more is emitted with a delay since the neutrinos must
diffuse out. Once the black hole has formed, the neutrino-emitting material falls within the event horizon much faster than the neutrino diffusion timescale, and any energy that has not been released by the time of collapse ends up in the black hole \citep{Pra97, Bur88}. The protoneutron star is also extremely hot and thus behaves in a slightly different manner from cold neutron stars. We represent the protoneutron star
as a low-entropy core surrounded by a hot envelope and track three
masses: $M_{Gh}, M_B,$ and $M_{th}$.

$M_B$ is the baryonic mass, i.e. all mass that either was inside the
inner boundary at the time of collapse or subsequently accretes through
the inner boundary.  $M_{Gh}$ is the effective gravitational mass of
the hot neutron star, which is \emph{not}, in general, equivalent to
$M_o$, the gravitational mass of a \emph{cold} neutron star with the
same baryon content. This is the mass that defines the gravitational
potential in the outer part of the star.  $M_{th}$ accounts for the
difference in mass-energy stored in this hot neutron star as opposed
to a cold one, due to both its higher internal energy and inflated
radius. This extra internal energy diffuses away on the cooling
timescale $\tau_c$. The gravitational mass $M_{Gh}$ of the hot neutron star
is then:
\begin{eqnarray}
M_{Gh} &= M_o + M_{th} \\
               &= M_B - B\!E_c + M_{th} \label{eq:Mth}
\end{eqnarray}
where $B\!E_c$ is the cold neutron star binding energy. As matter
accretes a fraction $\epsilon$ of the subsequent change in binding
energy is assumed to be trapped and added to $M_{th}$ while the
remaining fraction 1 - $\epsilon$ is radiated promptly as
neutrinos. The trapped energy diffuses out on a cooling
timescale $\tau_c$. The evolution of $M_{th}$ is thus given by 
\begin{equation}
\dot{M}_{th} = -\frac{M_{th}}{\tau_c} + \ep \, \frac{d \, \BE_c}{dM_B} \ \dot{M}_B
\end{equation}
and the initial condition $M_{th} = B\!E_c$. The derivative $d \,
\BE_c / dM_B$ is evaluated from the binding energy equation,
Eq. \ref{eq:BE}. Once all $M_{th}$ has diffused away, i.e. the neutron
star has cooled, Eq. \ref{eq:Mth} reduces to the standard cold neutron
star equation $M_G = M_B - \BE_c$. Assuming rapid virialization of the accreted material implies that the parameter $\epsilon$ should not exceed $\sim 0.5$. As the cooling timescale at the surface is likely less than the timescale for the full PNS, $\epsilon$ is likely to be much lower. In the limit that the cooling timescale were much shorter than the accretion timescale, $\epsilon$ would go effectively to zero. For our models we adopt $\epsilon = 0.1$. In Section \ref{sec:realistic} we test the effects of varying this parameter and find them to be small.

The emission halts when the neutron star has accreted past the TOV
maximum mass limit. However this limiting mass must be adjusted for the
neutron star's evolving heat content. Therefore we wait until the cold
core of the neutron star ($M_{Gh} - M_{th}$) has accreted past the TOV
limit before we assume the core has become a black hole and shut off
the neutrino emission. From this point on the central object behaves
as a purely gravitational point mass that absorbs all of
$\dot{M}_B$. This is referred to as the ``full loss'' model.

\begin{figure}
\includegraphics[scale = 0.5, trim = 20 10 0 25, clip = true]
{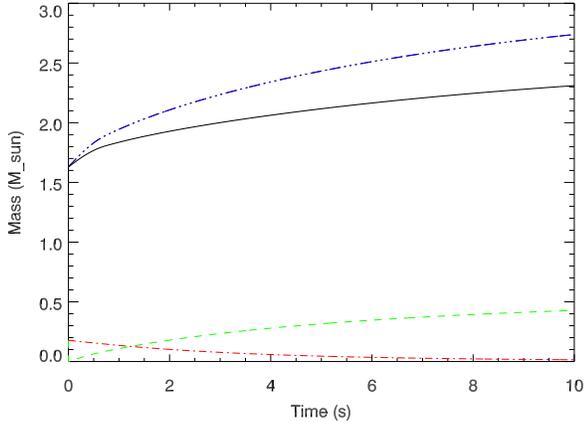}
\caption{Core mass growth in the 15 $\Msun$ model RSG15 with an
  assumed TOV mass limit of 2.5 $\Msun$, demonstrating the ``full
  loss'' model. The black curve (solid) shows $M_{Gh}$, the gravitational
  mass. Blue (dash-triple dot) is $M_B$, the baryonic mass. Red (dash-dot) is the thermal mass
  $M_{th}$. Green (dashed) shows the cumulative lost mass from the simulation.}
\label{fig:mass}
\end{figure}

Figure \ref{fig:mass} shows an example of the full model for
RSG15, TOV = 2.5 $\Msun$. The blue curve shows $M_B$, the total
baryonic mass that has entered the core region (precollapse core plus
accretion); it is higher than the black curve showing $M_{Gh}$, the
effective gravitational mass of the core. The red curve shows
$M_{th}$, which decays exponentially as the protoneutron star
cools. The green curve shows the cumulative mass loss in the
simulation. Even though the core in this case lives for almost 24 s as
a neutron star, most of the mass loss occurs during the first 5 seconds.


\subsection{Choice and Structure of Stellar Models}

\begin{figure}
\includegraphics[scale = 0.5, trim = 20 10 0 25, clip = true]{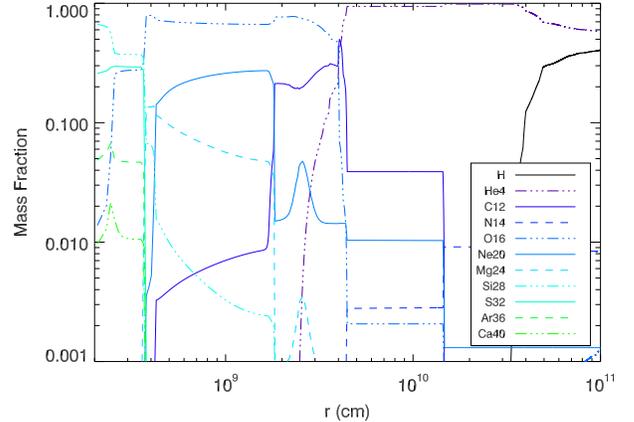}
\caption{Composition of the RSG15 supernova progenitor star.}
\label{fig:rsg15}
\end{figure}

\begin{figure}
\includegraphics[scale = 0.5, trim = 20 10 0 25, clip = true]{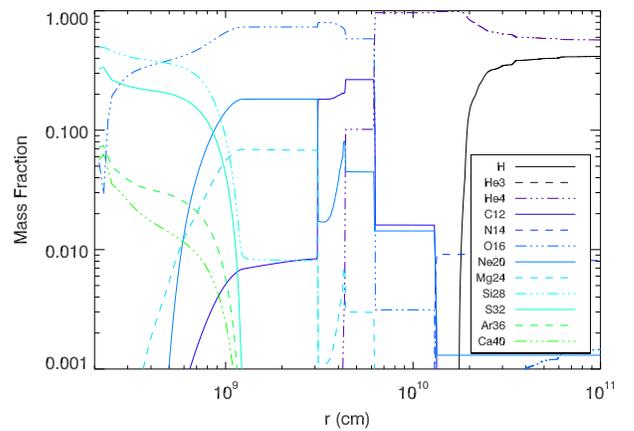}
\caption{Composition of the RSG25 supernova progenitor star.}
\label{fig:rsg25}
\end{figure}

The two presupernova models used for this paper were red supergiants
with a ZAMS mass of 15 $\Msun$ (RSG15), shown in Fig. \ref{fig:rsg15}, and 25 $\Msun$ (RSG25), shown in Fig. \ref{fig:rsg25}, both of
solar metallicity. They are taken from the survey of
\citet{Woo07}. Both stars shed substantial amounts of mass before
reaching the end of their lives.  The 15 $\Msun$ star represents a
more common supernova progenitor, while the 25 $\Msun$ model has a
more compact structure and may be more likely to make a black hole
promptly. The observed upper limit on the progenitor masses of standard core-collapse supernovae found
to date is 18 $\Msun$ \citep[e.g.][]{Sma09},
suggesting that higher mass progenitors are more likely to collapse in
a non-traditional way, or that these progenitors are preferentially obscured. Model RSG15 has a helium core of 4.27 $\Msun$
that extends to $3.568\E{10}$ cm. RSG25 has a helium core of 8.20
$\Msun$ that extends to $1.807\E{10}$ cm. Both red supergiants have an
extended, tenuously-bound hydrogen envelope extending from the helium
core out to $\sim 5\times10^{13}\textrm{ cm}$ that is easily
ejected. The net binding energy of the envelope external to what is
nominally the helium core (where the hydrogen mass fraction declines
to 1\%) is $1.1 \times 10^{48}$ erg, but a short distance out at 4.48
\Msun \ this declines to $10^{47}$ erg, a value that characterizes
most of the hydrogen envelope. These values include internal energy
but not the energy potentially available from recombination. For the
25 \Msun \ model, the net binding energy external to the helium core
is $6.4 \times 10^{48}$ erg, but this declines to 10$^{47}$ erg when
the interior mass increases from 8.20 \Msun \ to 8.36 \Msun and a
still smaller value characterizes most of the envelope. We note that
the recombination of 10 \Msun \ of hydrogen (13.6 eV per atom) would
release $2.6 \times 10^{47}$ erg.

Some additional parameters of both models can be found in Table 1. Also
given is the compactness parameter, $\xi_{2.5}$, as defined by
\citet{Oco11}, but computed at the time when the star is moved to
CASTRO rather than at the time of core bounce. A star with a higher
compactness parameter has a denser region surrounding its core; for
our purposes, this means it will accrete to the TOV limit faster and
therefore potentially lose less mass as neutrinos.


Because of the extended size of the red supergiant progenitors, we do
not carry the entire star on our simulation grid in CASTRO. Instead we
model only the helium core and base of the hydrogen envelope.
\begin{figure}
\includegraphics[scale = 0.55, trim = 35 0 0 25, clip = true]{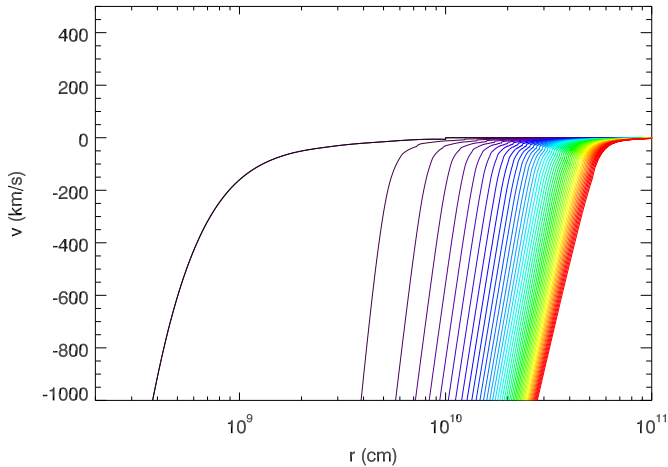}
\caption{Velocity curves showing collapse of the RSG15 stellar model in the prompt black hole
  formation case, i.e. without any mass loss. Positive velocities are outwards, negative are inwards. Curves are purple for early times, shading to red for late, approximately 15 s spacing. With no mass decrement and no core bounce shock to provide outward velocity, the star
  simply falls into the black hole. Total time shown: 709 s.}
\label{fig:nov}
\end{figure}

\begin{deluxetable*}{ccccc}
\tablewidth{0pt}
\tablecaption{Stellar Model Parameters}
\tablehead{
\colhead{Model Name} &
\colhead{Final Mass ($\Msun$)} &
\colhead{He Core Mass ($\Msun$)} &
\colhead{He Core Radius (cm)} &
\colhead{Compactness $\xi_{2.5}$}}
\startdata
RSG15 & 12.79 & 4.27 & 3.568\E{10} & 0.18\\
RSG25 & 15.84 & 8.20 & 1.807\E{10} & 0.33
\enddata
\end{deluxetable*}

\section{Hydrodynamic Response}
\subsection{Models Without Mass Loss}

Without an outgoing shock or mass loss from neutrinos, the inner
layers of the star should collapse directly into the black hole. This
scenario provides an excellent check of the fidelity of our
simulation. As Figure \ref{fig:nov} shows, our model accurately
reproduces this behavior. Dark (purple) colors on the plot indicate
early times, shading to light (red) colors at late times. Initially
the bulk of the star is in hydrostatic equilibrium (zero velocity)
with a small portion near the core showing high infall velocities. As
the collapse continues, more and more of the stellar material acquires
negative velocities. Without an outgoing shock to counter this motion,
it eventually falls into the core. If the simulation is run for long
enough, all mass will disappear from the grid. Physically, this
scenario corresponds to prompt black hole formation, when the core
collapses without passing through an intermediate protoneutron star
stage. If the outer layers of the star do not have sufficient angular
momentum to form a disk when they reach the black hole, the entire
star will disappear without producing a transient - the unnova case as
defined by \citet{Koc08}.

\begin{figure}
\includegraphics[scale = 0.55, trim = 35 0 0 0, clip = true]{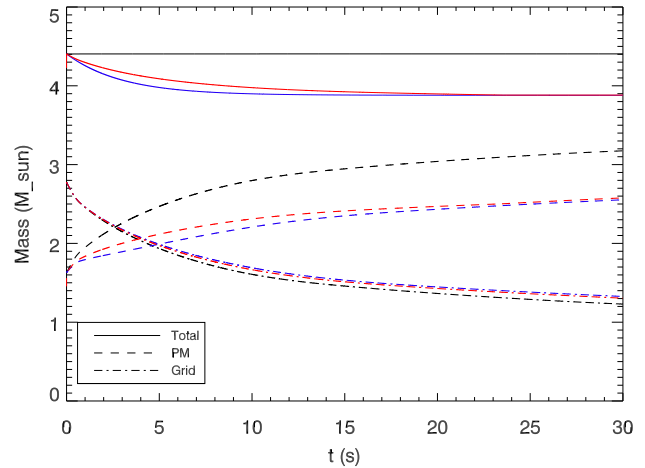}
\caption{Various core masses as a function of time for Model RSG15 and
  a TOV limit of 2.5 \Msun. The central mass $M_{Gh}$ (dashed) grows while the mass on the simulation grid
  (dot-dashed) drops. The solid line represents the total mass in the
  center and on the grid combined.  Black curves show the no neutrino
  loss (constant mass) case.  Blue curves show the maximum-loss model,
  while the red curves show the full neutrino loss model.}
\label{fig:pm}
\end{figure}

Figure \ref{fig:pm} shows the growth of the simulated point mass in RSG15. 
The black curves show the model with no neutrino mass loss; in this case the point mass corresponds to $M_B$. The
dashed line represents the growth of the point mass while the dot-dashed line shows
the mass on the simulated grid. Over time this point mass grows as the grid mass declines, with
most of the change occurring in the first 15 seconds. The solid line
shows the sum of the point mass and grid mass, i.e. the total mass
represented in the simulation; in the no-mass-loss case it is constant
throughout. This confirms that our simulation is reproducing the
collapse accurately, without spurious shocks or unphysical mass loss.

\subsection{Models With Maximum Mass Loss}

The set of blue curves in Figure \ref{fig:pm} shows the evolution
using the simplified maximum mass loss model. In this case, the point mass corresponds to $M_G$ with losses defined by Eq. \ref{eq:max}, in which the
core loses the total binding energy appropriate to a TOV-limit neutron star on a time scale $\tau_c$, regardless of the amount of mass
flowing in from the collapsing star. This model therefore provides an
upper bound on possible transients, as the core cannot lose more mass
than the binding energy of the largest possible neutron star. The
amount of mass lost in each case is listed in Table 2. Following the
blue curves in Figure \ref{fig:pm}, as the collapse begins the point
mass growth (dashed) is noticeably suppressed for the first 5 seconds
as neutrino losses balance accreted mass. The overall mass in the
simulation (solid) drops accordingly, then becomes constant as mass
loss ceases.

\begin{figure}
\includegraphics[scale = 0.55, trim = 35 0 0 25, clip =  true]{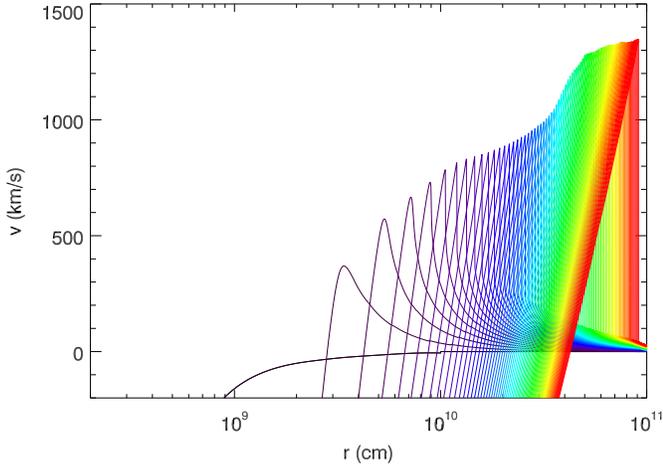}
\caption{Collapse of the RSG15 model in the maximum neutrino mass loss
  case, showing a shock forming due to the effective core mass
  decrement. Positive velocities are outwards, negative are inwards. Curves are purple for early times, shading to red for late. This model has the TOV limit set to 2.5, resulting in a
  mass decrement of 0.525 \Msun. The shock propagates out of the
  helium core ($r$ = 3.568\E{10} cm). Time shock reaches $10^{10}$ cm: 38 s. Time to end of helium core: 196 s. Time to $10^{11}$ cm: 577 s.}
  \label{fig:max}
\end{figure}

This mass decrement was sufficient to produce an outgoing shock in the
inner layers of the 15 \Msun \ presupernova star. The shock's
evolution is seen in Figure \ref{fig:max} (purple at early times, shading to red at late times). The shock
grows in speed as it leaves the helium core, and succeeds in reaching
the base of the hydrogen envelope. Of interest is the fact that the
shock strength varied noticeably with the choice of neutron star upper
mass limit. The approximate shock strengths at $10^{11}$ cm for our
six different choices of TOV limit are listed in Table 2.

\subsection{Models With Realistic Neutrino Mass Loss}
\label{sec:realistic}
The red curves in Figure \ref{fig:pm} show the mass evolution for the full model for neutrino
losses described in Section 2.1. In this case the dashed line showing the point mass corresponds to $M_{Gh}$ as given by Eq. \ref{eq:Mth}. This model takes into account the
thermal mass loss and ties the cessation of neutrino losses to the
amount of material that has been accreted by the core, rather than
switching it off after a predetermined timescale as in the upper bound
model. As can be seen in Figure \ref{fig:pm}, this model (red) loses
mass over a longer timescale than the maximum loss model (blue),
continuing until the point mass reaches the TOV limit, in this case
2.5 \Msun, after which the total mass becomes constant. We expect
equal or less mass loss in this case as compared to the maximum loss
model. In the case where the TOV limit is high enough that the neutron
star lives for longer than the cooling timescale, the core loses close
to the maximum possible amount of mass; in the case where it does not,
however, mass loss is suppressed as neutrinos that would have been emitted instead end up inside the black hole. The amount of mass lost in each case
is listed in Table 3.
\begin{figure}
\includegraphics[scale = 0.55, trim = 35 0 0 25, clip = true]{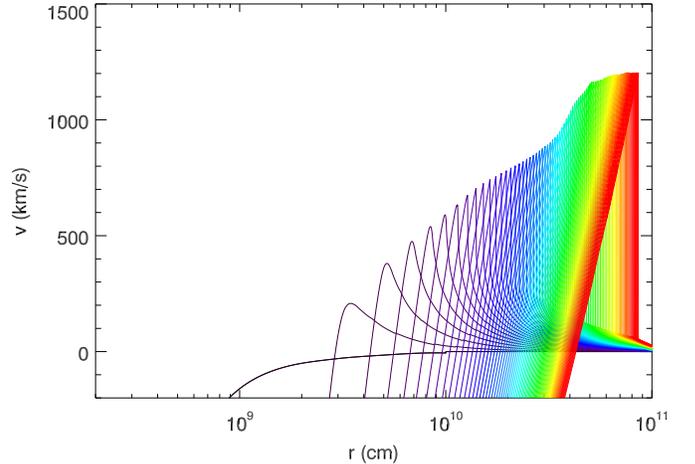}
\caption{Collapse of the RSG15 model in the fully-modeled neutrino
  mass loss case, showing a shock forming due to the effective core
  mass decrement. Positive velocities are outwards, negative are inwards. Curves are purple for early times, shading to red for late. This model has the TOV limit set to 2.5, resulting
  in a mass decrement of 0.523 \Msun. The shock is smaller in strength
  than the maximum-loss case and reaches the edge of the simulation
  with a lower velocity. Time shock reaches $10^{10}$ cm: 40 s. Time to end of helium core: 207 s. Time to $10^{11}$ cm: 620 s.}
  \label{fig:full}
\end{figure}

\begin{figure}
\includegraphics[scale = 0.55, trim = 35 0 0 25, clip =
  true]{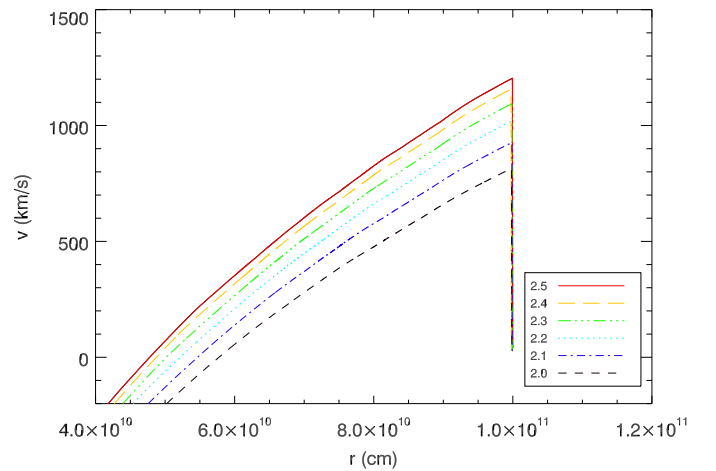}
\caption{RSG15 shocks at the limit of the CASTRO simulated domain for six
  different choices of TOV limit, full neutrino loss model.}
\label{fig:rsg15lim}
\end{figure}

Though the overall mass decrement in the full model cases is lower
than in the maximum loss case, it is still sufficient to produce an
outgoing shock. Figure \ref{fig:full} shows the shock evolution for RSG15, TOV =
2.5 $\Msun$. The approximate shock strengths at $10^{11}$ cm for our
six different choices of TOV limit are listed in Table 3. The six
shocks created in RSG15 are shown in Figure \ref{fig:rsg15lim}.

\begin{figure}
\includegraphics[scale = 0.55, trim = 35 0 0 25, clip = true]{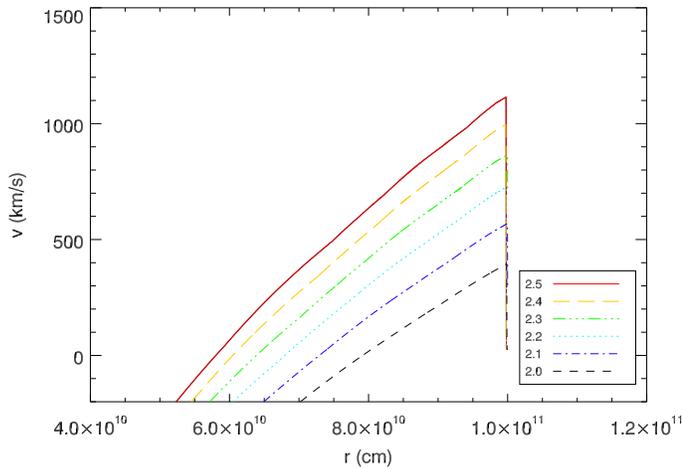}
\caption{RSG25 shocks at the limit of the CASTRO simulated domain for six
  different choices of TOV limit, full neutrino model. The choice of
  TOV limit has a stronger effect on the final shock strength in this model than it does in RSG15.}
\label{fig:rsg25lim}
\end{figure}

\begin{deluxetable*}{ccccc}
\tablewidth{0pt}
\tablecaption{Maximum Mass Loss Models}
\tablehead{
\colhead{Stellar Model} &
\colhead{TOV ($\Msun$)} &
\colhead{Mass Lost ($\Msun$)} &
\colhead{KE\tablenotemark{a} (ergs)} &
\colhead{Shock Strength\tablenotemark{b} (km/s)}}
\startdata
RSG15 & 2.0 & 0.336 & 1.875\E{47}& 902\\
\nodata & 2.1 & 0.370 & 2.599\E{47}& 985\\
\nodata & 2.2 & 0.407& 3.572\E{47}& 1070\\
\nodata & 2.3 & 0.444 & 4.855\E{47}& 1158\\
\nodata & 2.4 &  0.484 & 6.554\E{47}& 1249\\
\nodata & 2.5 & 0.525 & 8.719\E{47}& 1341\\
\hline \\
RSG25 & 2.0 & 0.336 & 6.537\E{46} & 723\\
\nodata & 2.1 & 0.370 & 1.002\E{47} & 820\\
\nodata & 2.2 & 0.407 & 1.483\E{47} & 919\\
\nodata & 2.3 & 0.444 & 2.139\E{47} & 1025\\
\nodata & 2.4 & 0.484 & 3.011\E{47} & 1134\\
\nodata & 2.5 & 0.525 & 4.148\E{47} & 1246
\enddata
\tablenotetext{a}{At base of hydrogen envelope}
\tablenotetext{b}{At $r$ = 10$^{11}$ cm}
\end{deluxetable*}

\begin{deluxetable*}{ccccc}
\tablewidth{0pt}
\tablecaption{Full Mass Loss Models}
\tablehead{
\colhead{Stellar Model} &
\colhead{TOV ($\Msun$)} &
\colhead{Mass Lost ($\Msun$)} &
\colhead{KE\tablenotemark{a} (ergs)} &
\colhead{Shock Strength\tablenotemark{b} (km/s)}}
\startdata
RSG15 & 2.0 & 0.277 & 1.287\E{47} & 814\\
\nodata & 2.1 & 0.331& 2.059\E{47}& 926\\
\nodata & 2.2 & 0.382 & 2.953\E{47}& 1019\\
\nodata & 2.3 & 0.430 & 3.911\E{47}& 1094\\
\nodata & 2.4 & 0.477 & 4.896\E{47}& 1157\\
\nodata & 2.5 & 0.523 & 5.779\E{47}& 1204\\
\cline{1-5}\\
RSG25 & 2.0 & 0.179 & 8.418\E{45} & 394	\\
\nodata & 2.1 & 0.230 & 2.893\E{46} & 569\\
\nodata & 2.2 & 0.281 & 6.581\E{46} & 725\\
\nodata & 2.3 & 0.331 & 1.204\E{47} & 866\\
\nodata & 2.4 & 0.382 & 1.930\E{47} & 996\\
\nodata & 2.5 & 0.433 & 2.827\E{47} & 1114
\enddata
\tablenotetext{a}{At base of hydrogen envelope}
\tablenotetext{b}{At $r$ = 10$^{11}$ cm}
\end{deluxetable*}

We also tested variations in the parameter $\epsilon$, which controls the fraction of binding energy trapped as thermal mass. Changes in $\epsilon$ have a small but real effect on the total mass loss, depending on the amount of accreted mass. The more mass accreted, the more important $\epsilon$ will be. As higher TOV limit models tend to accrete longer, $\epsilon$ has a higher impact here. A lower epsilon leads to a higher mass loss as less of the binding energy is temporarily trapped as thermal mass. We tested the range $\epsilon < 0.5$, identified as the physically reasonable range of this parameter. For the case where the TOV limit is 2.5 $\Msun$, the most sensitive, a change of 0.05 in $\epsilon$ in RSG25 resulted in approximately a 0.011 $\Msun$ change in the overall mass loss. In the extreme case $\epsilon = 0.5$, this will make a TOV = 2.5 $\Msun$ model look like a TOV $\sim$ 2.35 $\Msun$ model. For lower TOV limits, the effect of varying $\epsilon$ was smaller. Thus, although a change in $\epsilon$ will affect the mass decrement and by extension the shock strength, the results are robust.

\subsection{Effects of Stellar Structure}

The kinetic energy of the shock that reaches the hydrogen envelope is
strongly influenced by the size of the presupernova core and the overlying stellar structure through which
it must travel, as can be seen by the differences in kinetic energy
between RSG25 and RSG15 in the maximum mass loss model. This model
loses the same amount of mass in the same time regardless of stellar
structure, but the final shock in RSG25 models is significantly weaker
than in RSG15, having propagated through a much heavier carbon-oxygen and helium
core before reaching the hydrogen envelope.

Additional effects come into play when using the full neutrino model
based on the speed at which the core accretes. RSG25 has a denser
inner structure (higher compactness) and a more massive iron core of
1.83 $\Msun$, compared to 1.63 \Msun. As it therefore accretes faster
than RSG15 and is already closer to the TOV limit, the core spends
significantly less time in the protoneutron star state; consequently
the same choice of parameters in the larger star causes less mass
loss. The full mass loss model in RSG25 also shows systematically
lower mass decrements than the maximum mass loss model in all cases,
indicating that even with a high TOV limit and a relatively long-lived
protoneutron star, not all the binding energy is being emitted before collapse to a
black hole. The six shocks produced in RSG25 are shown in Figure
\ref{fig:rsg25lim}.

\section{Transients Produced}

\begin{figure}
\includegraphics[scale = 0.55, trim = 35 0 0 0, clip =
  true]{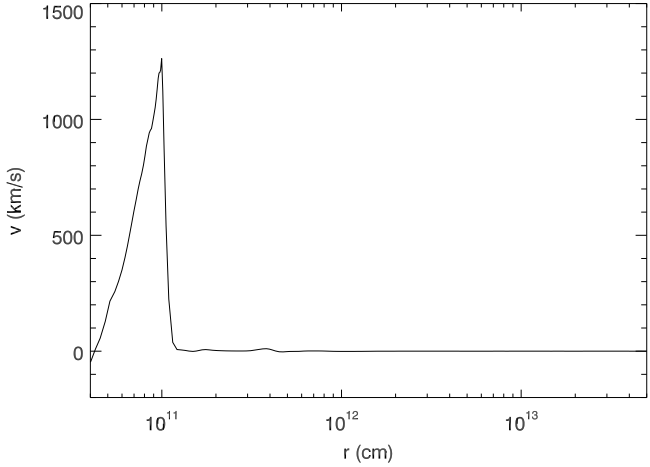}
\caption{Shock modeled by CASTRO in RSG15 (TOV = 2.5, full neutrino losses) mapped into KEPLER.}
\label{fig:keplershock}
\end{figure}

\begin{figure}
\includegraphics[scale = 0.55, trim = 35 0 0 0, clip =
  true]{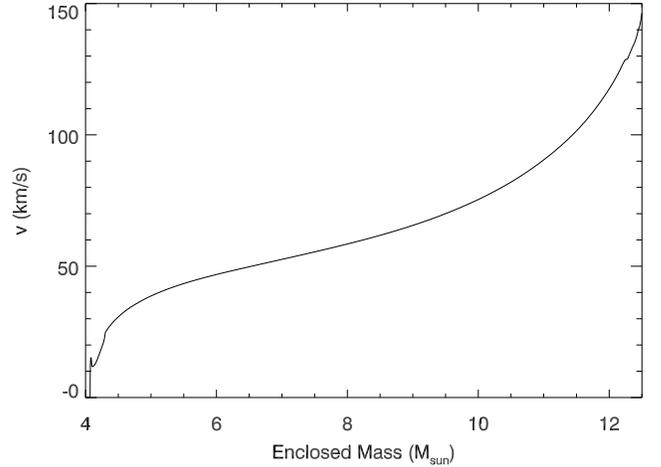}
\caption{Velocity of the hydrogen envelope at 5\E{7} s after core
  collapse in RSG15, TOV = 2.5, full neutrino loss model, evolved further in KEPLER.}
\label{fig:keplerenv}
\end{figure}

In the case of prompt black hole formation in a star without a
high-$J$ outer layer, there is no shock and no visible transient - the star simply
disappears. Here we consider further the case of protoneutron
star formation and delayed collapse, in which an outgoing shock has formed. Will a detectable transient be
produced, intermediate between a complete disappearance and a traditional explosion, as raised by \citet{Koc08}? To evaluate this question we took models from CASTRO where
the shock had reached $1\E{11}$ cm (the limit of the simulation) and
mapped them back into KEPLER, then continued to evolve them. Figures
\ref{fig:keplershock} and \ref{fig:keplerenv} show the KEPLER results
for RSG15, TOV = 2.5; Figure \ref{fig:keplershock} shows the imported
shock, and Figure \ref{fig:keplerenv} shows the final velocity of the
hydrogen envelope at $t = 5\E{7}$ s.

\begin{figure}
\includegraphics[scale = 0.55, trim = 20 0 0 0, clip =
  true]{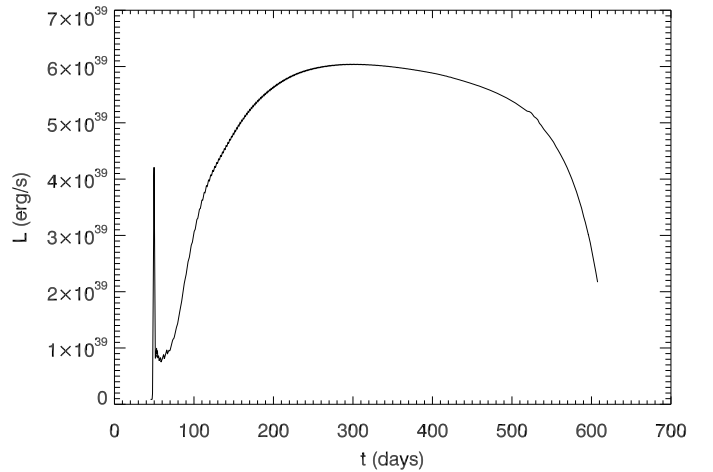}
\caption{KEPLER light curve for a transient from RSG15, TOV = 2.5. The
  transient is low luminosity but lasts for around a year.}
\label{fig:keplerlc}
\end{figure}

The shock has decreased significantly in strength by the time it
reaches the base of the hydrogen envelope; however, this envelope is
very tenuously bound in both RSG15 and RSG25. For each model we tested
six choices of TOV limit (2.0 - 2.5 $\Msun$, in 0.1 $\Msun$
increments) and evaluated the strength of the shock that reached the
hydrogen envelope. Using the full neutrino loss model, we found in
every case tested for RSG15 and in 3 of 6 tested for RSG25 that the
shock produced was larger than 1\E{47} ergs, the approximate binding
energy of the envelope (see Table 3). We can therefore realistically
expect the envelope to be ejected in these cases. However, the highest
kinetic energy achieved was only of the order of 6\E{47} ergs, and
most models fell well below that number. The envelope is therefore
ejected with a very low velocity (50 - 100 km/s). It emits most of its
energy via hydrogen recombination. Optically this transient has a low
luminosity $\sim 10^{39} - 10^{40}$ ergs/s, but maintains this
luminosity for of order a year. The color temperature of the transient
is very red, of order 3000 K. An example light curve can be seen in
Figure \ref{fig:keplerlc} for RSG15, TOV = 2.5.

The transients calculated here are obviously much fainter and less
energetic than standard core-collapse supernova, but they do bear some
similarity to a class of recently-observed transients: the ``luminous
red novae," such as V838 Mon \citep{v838}. Luminous red novae are too
bright to be ordinary classical novae, but too faint and red to be
supernovae. Although V838 Mon is now suspected to be a stellar merger event \citep{Tyl11}, these two mechanisms have similar end results: a massive hydrogen envelope ejected at low energies. Spectroscopic observations show however that these
phenomena have dispersion velocities significantly higher than calculated
here. The observation of further transients may decide this question,
or a search for remnants. The shedding of a common envelope by a
binary merger would leave behind a degenerate remnant, but a failed
core-collapse explosion would leave a black hole. A survey such as that proposed by 
\citet{Koc08}, monitoring red supergiants for anomalous transients that might signal the birth of a black hole, should catch these events. They would be visible as a sudden brightening of the ``star" for of order a year, followed by a gradual but complete disappearance.

In RSG25 a TOV limit of 2.2 $\Msun$ or lower resulted in such weak
outgoing shocks that they could not be accurately followed using
KEPLER, and would probably be unable to eject the envelope. In
situations where the envelope is not ejected, there is still the
possibility of a transient at late times if the envelope is
rotating. As it falls back into the black hole, the massive envelope
may create a disk and potentially a long-duration gamma-ray transient
as described by \citet{Woo12}. Since the most massive stars are the
ones more likely to produce black holes quickly, it remains possible
to produce these long gamma-ray transients. This type of transient,
while invisible in the optical, could emit low levels of gamma rays
for months.

A higher TOV limit in the neutron star EOS will increase the probability of these transients occurring. Holding the TOV limit constant, the final strength of the shock is highest in stars with both smaller initial iron core masses (more time spent as a neutron star) and smaller carbon-oxygen and helium core masses. We might therefore expect the strongest transients to come from the lowest-mass red supergiants that fail to form CCSNe. Nucleosynthetic constraints place a lower limit on the maximum mass star that must explode as a supernova most of the time. \citet{Bro13} sets this limit at between 20 and 25 $\Msun$. Stars above 20 $\Msun$ become more difficult to explode, as measured by their compactness parameter \citep{Oco11}, and hence more likely to fail. At the same time, in stars above 25 $\Msun$, it will become increasingly difficult for the shock to reach the surface. We may therefore expect the progenitors of these transients, if they do occur, to land in the range 20 - 25 $\Msun$.

For heavier stars that lose their hydrogen envelope and die as WR stars, or for stripped progenitors in binaries, a shock can form and may reach the surface, depending on the size of the remaining helium core. Without a large envelope to eject, the transient will be brief but brighter.

\section{Conclusions}

We have demonstrated that iron core collapse in a massive star is
capable of producing a faint observable transient even if the collapse
itself creates no prompt outgoing shock.  The mass lost to neutrinos
results, in some cases, in unbinding the hydrogen envelope. The amount
and history of the neutrino mass loss has a strong effect on the
magnitude of the shock produced, as does the structure of the
carbon-oxygen and helium cores of the progenitor star. In the two red
supergiant models tested, the shock reached the base of the hydrogen
envelope in a majority of the models with enough energy to eject it. These
unusual transients will appear as low-energy, long-duration, red
events as the ejected envelope emits its energy via hydrogen
recombination. The ejected envelope has a speed on the order of 50 -
100 km/s and maintains a luminosity $10^{39} - 10^{40}$ ergs/s for
approximately a year.

For a given parametrization of the neutrino losses, the transient
produced becomes weaker as the TOV mass limit is reduced and as the
mass of the presupernova helium core increases. It therefore remains
possible, depending upon the TOV limit assumed, to fail to eject the
envelope in more massive stars. If the star in these cases has sufficient angular momentum
in its outer layers, it may instead produce long gamma-ray
transients as described by \citet{Woo12}; otherwise it will disappear as
an unnova as described by \citet{Koc08}.

\section{Acknowledgements}

We thank Luke Roberts for insight and guidance in modeling the
neutrino losses from a protoneutron star. We thank Ann Almgren and
John Bell for providing the CASTRO code and Shawfeng Dong, Chris
Malone, and Haitao Ma for assistance with using it. We thank Chris Kochanek and the referee for providing helpful comments.
 This research has been supported by the National Science Foundation (AST 0909129), the
NASA Theory Program (NNX09AK36G), the DOE High Energy Physics Program
(DE-FC02-09ER41438) and two UC Lab Fees Research Awards
(09-IR-07-117968 and 12-LR-237070).

\newpage


\clearpage

\end{document}